\documentclass[preprint]{aastex}
\def\fwidth{0.75\textwidth}

\newcommand{\lya}{Ly$\alpha$\ }

\newcommand{\psec}{\ensuremath{\, {\rm s}^{-1}}}

\newcommand{\hr}{\ensuremath{{\rm hr}}}

\newcommand{\m}{\ensuremath{{\rm m}}}

\newcommand{\Mpc}{\ensuremath{{\rm Mpc}}}
\newcommand{\K}{\ensuremath{{\rm K}}}

\newcommand{\mJy}{\ensuremath{{\rm mJy}}}

\newcommand{\ergs}{\ensuremath{{\rm ergs}}}

\newcommand{\Hz}{\ensuremath{\, {\rm Hz}}}
\newcommand{\kHz}{\ensuremath{\, {\rm kHz}}}
\newcommand{\MHz}{\ensuremath{\, {\rm MHz}}}

\begin{document}

\title{The 21 cm Forest as a Probe of the Reionization and
the Temperature of the Intergalactic Medium}

\author{Yidong Xu$^{1,2}$, Xuelei Chen$^{2}$, Zuhui Fan$^{1}$, Hy Trac$^{3}$,
 and Renyue Cen$^{4}$}
\altaffiltext{1}{Department of Astronomy, School of Physics, Peking
University, Beijing 100871, China} \altaffiltext{2}{National
Astronomical Observatory of China, CAS, Beijing 100012, China}
\altaffiltext{3}{Harvard-Smithsonian Center for Astrophysics,
Harvard University, 60 Garden St., Cambridge MA, 02138, USA}
\altaffiltext{4}{Department of Astrophysical Sciences, Princeton
University, Princeton, New Jersey 08544, USA}

\begin{abstract}
Using high redshift radio sources as background, the 21 cm forest
observations probe the neutral hydrogen absorption signatures of
early structures along the lines of sight. Directly sensitive to the
spin temperature of the hydrogen atoms, it complements the 21 cm
tomography observations, and provides information on the temperature
as well as the ionization state of the intergalactic medium (IGM).
We use a radiative transfer simulation to investigate the 21 cm
forest signals during the epoch of reionization. We first confirmed
that the optical depth and equivalent width (EW) are good
representations of the ionization and thermal state of the IGM. The
features selected by their relative optical depth are excellent
tracers of ionization fields, and the features selected by their
absolute optical depth are very sensitive to the IGM temperature, so
the IGM temperature information could potentially be extracted from
21 cm forest observation, thus breaking a degeneracy in 21 cm
tomographic observation. With the EW statistics, we predict some
observational consequences for 21 cm forest. From the distributions
of EWs and the number evolution of absorbers and leakers with
different EWs, we see clearly the cosmological evolution of
ionization state of the IGM. The number density of potentially
observable features decreases rapidly with increasing gas
temperature. The sensitivity of the proposed EW statistic to the IGM
temperature makes it a unique and potentially powerful probe of
reionization. Missing small-scale structures, such as small
filaments and minihalos that are unresolved in our current
simulation, and lack of an accurate calculation of the IGM
temperature, however, likely have rendered the presented signals
quantitatively inaccurate. Finally we discuss the requirements of
the background radio sources for such observations, and find that
signals with equivalent widths larger than 1kHz are hopeful to be
detected.

\end{abstract}


\section{Introduction}

The reionization of the intergalactic medium (IGM) is one of the
landmark events in the history of the universe.
While a lot of theoretical works
\citep[e.g.]{WL03,FZH04a,ZHH07,CF07,Choudhury09} and observational
constraints \citep[e.g.]{Fan06,Dunkley08,Gallerani08} have schemed
up a global picture of the reionization process, many detailed
issues remain uncertain \citep{CF05}. One of the most promising
probes of the cosmic reionization is the redshifted 21 cm transition
of HI (see e.g. \citealt{FOB06} for a review). The emission or
absorption of the 21 cm radiation is directly related to the neutral
component of the IGM at different redshifts, thus affording us the
most clear view of the reionization history. Moreover, unlike the
Ly$\alpha$ resonance line which also traces the neutral hydrogen,
the 21 cm line has an extremely small Einstein coefficient ($A_{10}
= 2.85 \times 10^{-15}s^{-1}$), hence it is not saturated even at
high redshift, and one can therefore safely predict the neutral
fraction from its 21 cm optical depth.

The most commonly discussed way of 21 cm observation is {\it 21 cm
tomography}, in which  the three dimensional structure of the
emission or absorption of 21 cm photons by the IGM against the CMB
is observed \citep{MMR97,Tozzi00,Iliev02,FZH04a,FZH04b}. A less
frequently discussed way of 21 cm observation is to observe the
spectra of high redshift radio point sources, such as quasars or
gamma ray burst (GRB) radio afterglows, and look for absorption or
leak features produced by structures along the line of sight. Such
{\it 21 cm forest} could be produced by structures of different
scales, including the large scale HI/HII regions (``bubbles''), the
cosmic web, proto-galaxies or mini-halos. The optical depth of the
21 cm line due to the absorption by the IGM at redshift $z$ is
\citep{Field59,MMR97,FOB06}:
\begin{eqnarray}
\tau_{\nu_0} (z) = \frac{3}{32\pi} \frac{h_p c^3 A_{10}}{k_B
\nu_0^2} \frac{x_{HI} n_H(z)}{T_S (1+z)(dv_{\parallel}/dr_\|)}
\\
\approx 0.009 (1+\delta) (1+z)^{3/2} \frac{x_{HI}}{T_S} \left[
\frac{H(z)/(1+z)}{dv_{\parallel}/dr_\|} \right],\nonumber
\end{eqnarray}
where $x_{HI}$ is the neutral fraction of hydrogen at a specific
position, $n_H(z)$ and $T_S$ are the number density of hydrogen and
the spin temperature of the IGM at that position, respectively, and
$dv_\|/dr_\|$ is the gradient of the proper velocity along the line
of sight, including both the Hubble expansion and the peculiar
velocity. As we are going to study
the large scale HI/HII regions, the peculiar velocity is neglected.
In the second line, the factor $(1+\delta)$ is the
overdensity of the baryons, and $T_S$ is in units of Kelvin.

In the following, we will assume that the spin temperature always follows the
gas temperature, i.e. $T_S = T_K$. This is a good approximation for the
redshifts relevant here \citep{Santos08}, where the
\lya background from star formations \citep{MMR97,CM04} is present, and
the HI spin temperature is strongly coupled to the kinetic
temperature via the Wouthuysen-Field effect \citep{W52,Field59}. The 21 cm forest
signal is sensitive to the spin temperature, this is a great advantage compared
with 21 cm tomographic observation against the CMB background. The latter could
also be sensitive to the spin temperature in the case of absorption, but once the
spin temperature raises above the CMB temperature, the 21 cm signal become emission,
which saturates and become almost independent of $T_S$. This uncertainty in
spin temperature causes large degeneracies between cosmological and
astrophysical parameters in 21 cm tomography \citep{SC06}.
The information extracted from the 21 cm forest observation may help
us to determine the spin temperature at the relevant redshifts, and
reduce the degeneracies in 21 cm tomography.

\cite{Carilli02} used numerical simulation to illustrate the 21 cm
forest imprinted on quasar spectrum. \cite{FL02} considered small
scale features produced by mini-halos and galaxies. The absorption
features produced by the HII regions (``ionized bubbles'') before
reionization was studied \citep{F06,X06} using the analytical
``bubble model'' of reionization \citep{FZH04a,FZH04b}, in which the
ionization proceeds inside-out, i.e. from high density regions to
low density regions.

The reionization process is very complicated, although the ``bubble
model'' picture of reionization is in general agreement with
simulation results \citep{Zahn07,Iliev07,Mellema06,TC07,STC08}, the
complicated geometry of the ionized regions is not considered in
this simple analytical model. Therefore, to extract information from
future 21 cm forest observations, it is imperative to examine how
the 21 cm forest observables are correlated with the reionization
process by using numerical simulations. In this paper, we study the
large scale 21 cm forest produced by the neutral and ionized bubbles
using the reionization simulation by \cite{STC08}. The high
resolution simulation has detailed modeling of star formation
and radiative transfer of ionizing photons. First, we check whether the equivalent width
(EW) can be a representative indicator of the ionization and thermal
state of the IGM, and to what extent the 21 cm forests trace cosmic
reionization. Then we use simulation data to predict the
observational consequences for 21 cm forest. Comparing them with
realistic parameters of radio arrays, we study what constraints we
could put on reionization from 21 cm forest observations.

We briefly describe the reionization simulation in section 2. In
section 3, we check whether or not, and to what extent, the optical
depth (hence the equivalent width) follows the ionization state
$x_i$, density field $\rho$, and thermal history
$T_{\ensuremath{{\rm IGM}}}$ of the IGM. From these analyses, we
would see what information can be extracted from EW statistics. Then
we predict the 21 cm forest signals -- distributions and evolution
of EWs -- from the reionization simulation in section 4. In section
5 we investigate whether there are sufficient number of background sources
that are luminous enough for such 21 cm forest observations. We
summarize and discuss our results in section 6.

Throughout the paper we use the same $\Lambda$CDM cosmology as
\citet{STC08} based on the WMAP3 results (see \citealt{Spergel07},
and references therein): $(\Omega_m, \Omega_\Lambda, \Omega_b, h,
\sigma_8, n_s) = (0.26, 0.74, 0.044, 0.72, 0.77, 0.95)$.

\section{The Simulation}

In this paper, we use the reionization simulation described in
detail in \cite{STC08}. The high-resolution simulation was run using
a hybrid code that contains a N-body algorithm for dark matter,
prescriptions for baryons and star formation, and a radiative
transfer algorithm for ionizing photons. All components are solved
simultaneously in a periodic box with comoving side length of 100
Mpc/h.

The N-body calculations included $2880^3$ collisionless dark matter
particles, each with a mass of $3.02\times10^6\ M_\odot/h$. This
provided sufficient resolution to resolve halos down to masses of
$\sim10^8\ M_\odot/h$ and account for the majority of photo-ionizing
sources. For each particle, the local matter density and velocity
dispersion are calculated in order to model the baryons and star
formation. The baryons are assumed to have the same overdensity
field as the dark matter, and the temperature field is set by the
velocity dispersion of the particles. Radiative heating and cooling
of the gas are not included in the hybrid simulation.

The criterion for star formation was that the particles must have
densities $\rho_m > 100 \rho_{crit}(z)$ and temperatures $T > 10^4$
K. This cut in the temperature-density phase space effectively
restricts star formation to regions within the virial radius of
halos that cool efficiently through atomic line transitions. The
simulation distinguishes between the first generation, Population
III (Pop III) stars and the second generation, Population II (Pop
II) stars. The stellar initial mass function is determined by
following the metallicity enrichment of the halo and the IGM as
described in \citet{TC07}. The input spectrum for ionizing radiation
is divided in three energy ranges (eV): $13.61\le\nu < 24.59$,
$24.59\le\nu <54.42$, and $\nu\geq 54.42$, according to the
ionization limits for for HI, HeI, and HeII.

The radiative transfer algorithm for ionizing radiation is based on
a photon-advection scheme that is computationally much less
expensive than the ray-tracing scheme in \citet{TC07}. These two
approaches have been demonstrated to give very similar results for
the majority of the reionization epoch \citep{STC08}. In the hybrid
simulation, reionization is completed by $z\approx 6$. For
post-analysis, the dark matter, baryons, and radiation are collected
on a grid with $720^3$ cells and the data is saved every 10 million
years from $z=25$ down to $z=5$. In this paper, we make use of the
gas density, ionization, and temperature fields.

\begin{figure}[tb]
\begin{center}
\includegraphics[width=\fwidth]{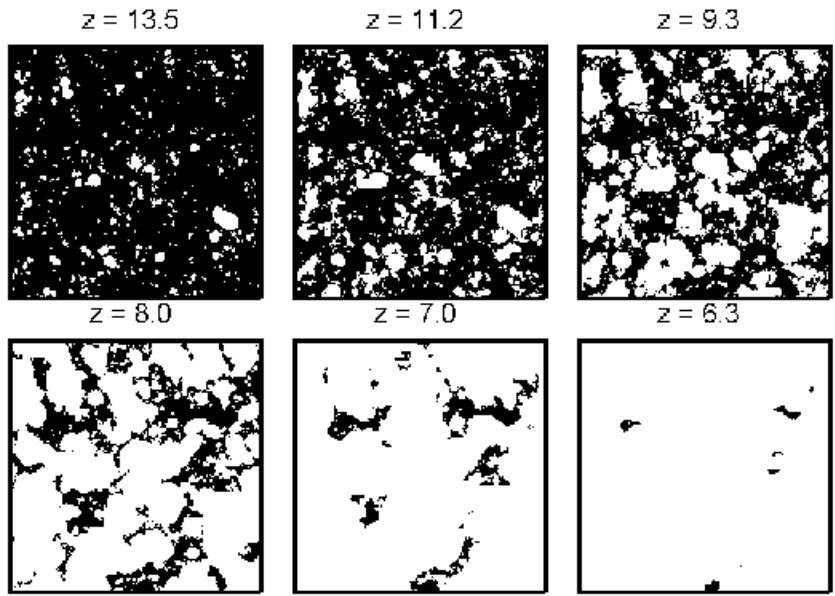}
\caption{\label{fig:xislice}Distributions of ionization fraction at
redshifts $z \sim$
13.5, 11.2, 9.3, 8.0, 7.0, and 6.3. Highly ionized regions are
represented by white while regions with ionization fraction below 50
percent is shown as black. These are plots of slices in the
simulation box with each side of 100 Mpc/h. The global ionization
fractions are $\sim $ 0.094, 0.24, 0.46, 0.73, 0.91,
and 0.98 respectively.}
\end{center}

\end{figure}

We mark a cell as ``ionized'' if the ionization fraction is greater
than 0.5, and ``neutral'' otherwise. Slices of ionization field for
several redshifts are shown in Fig.~\ref{fig:xislice}. Each slice in
the figure has a thickness of 139 kpc/h, and this applies to all the
other slice-plots in the following. We see that the HII bubbles were
highly biased and non-spherical; the universe became 50\% ionized at
redshift $z\sim 9.5$, and the non-gaussianity in the HII field was
significant throughout the reionization. In addition, the simulation
also shows that Pop III galaxies played a very important role in the
process, and a strong correlation existed between halo number
density and bubble size for large bubbles.

\begin{figure}[htb]
\begin{center}
\includegraphics[width=\fwidth]{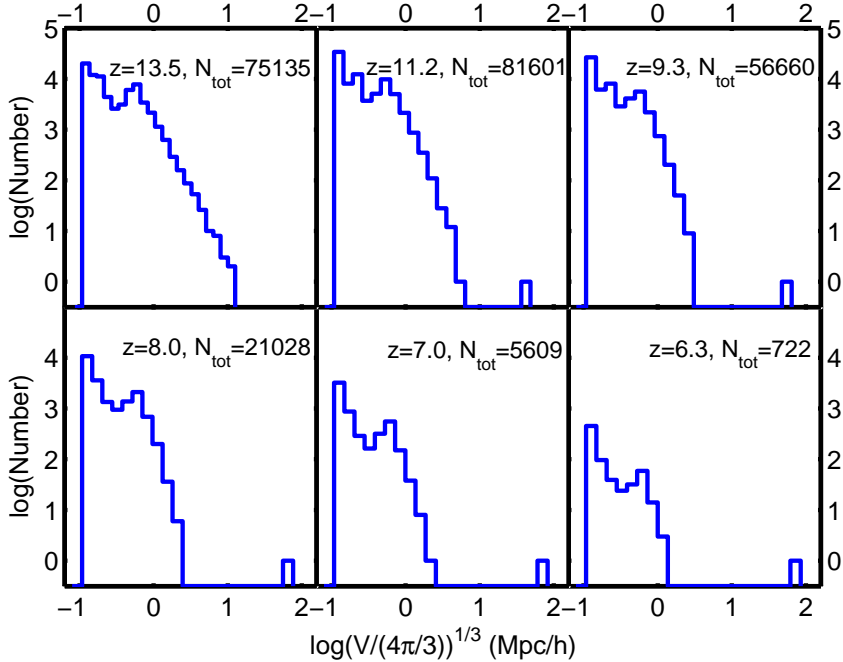}
\caption{\label{fig:xiBdist}Number distributions of $x_i$-bubble volumes at redshifts
$z =$ 13.5, 11.2, 9.3, 8.0, 7.0, and 6.3, respectively. }
\end{center}

\end{figure}

To characterize the ionized region statistically, we use the
Friend-of-Friend (FoF) algorithm to select ionized bubbles
(introduced by \cite{Iliev06}), which are made of connected ionized
cells. The linking length is set to be 1.2 times the simulation data
output cell size, so that cells are connected by one face at least.
In addition, one bubble consists of 2 or more connected cells.
Fig.~\ref{fig:xiBdist} shows the size distributions of bubbles
selected by ionized fraction ($x_i$-bubbles), with $x_i \ge 0.5$, at
several redshifts. As discovered by  \citet{STC08}, before the
percolation of the ionization bubbles, these
ionized-fraction-selected bubbles ($x_i$-bubbles) have the
distributions with three distinct peaks. The characteristic volumes
for the three peaks are 0.6, 0.03, and 0.006 Mpc$^3$/h$^3$
respectively. These sizes do not change significantly with
decreasing redshift, before one connected network of bubbles
dominated at $z \le 10$. The smallest peak in each distribution is
due to limited resolution of the simulation (i.e. cell size). The
other two characteristic sizes are believed to be real features of
the size distribution \citep{STC08}.

\section{The Equivalent Width as a Probe of Reionization}

\begin{figure}[htb]
\begin{center}
\includegraphics[width=\fwidth]{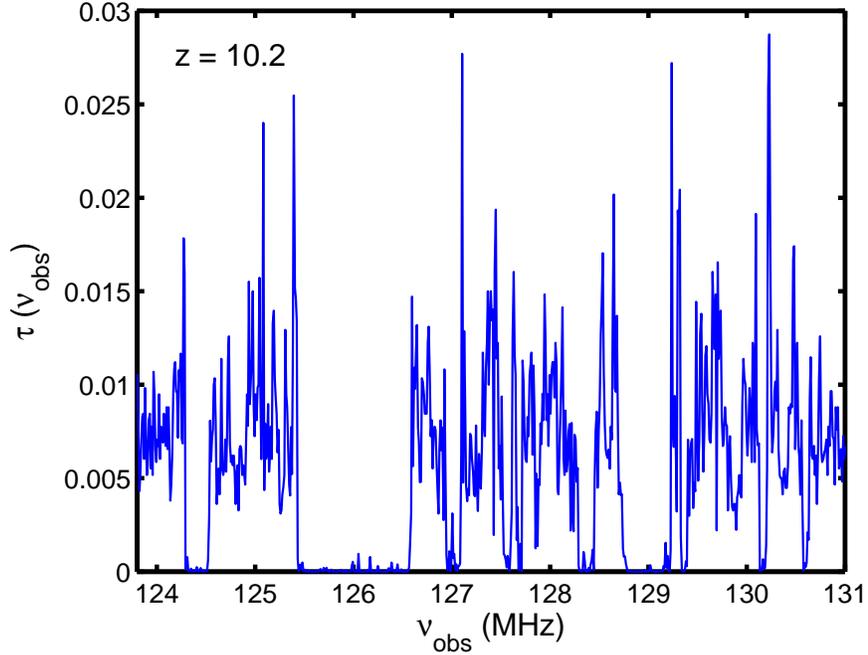}
\caption{\label{fig:example}The optical depth as a function of observed frequency along
a line of sight passing through the simulation box. The redshift
here is 10.2.}
\end{center}

\end{figure}

\begin{figure}[t]
\begin{center}
\includegraphics[width=\fwidth]{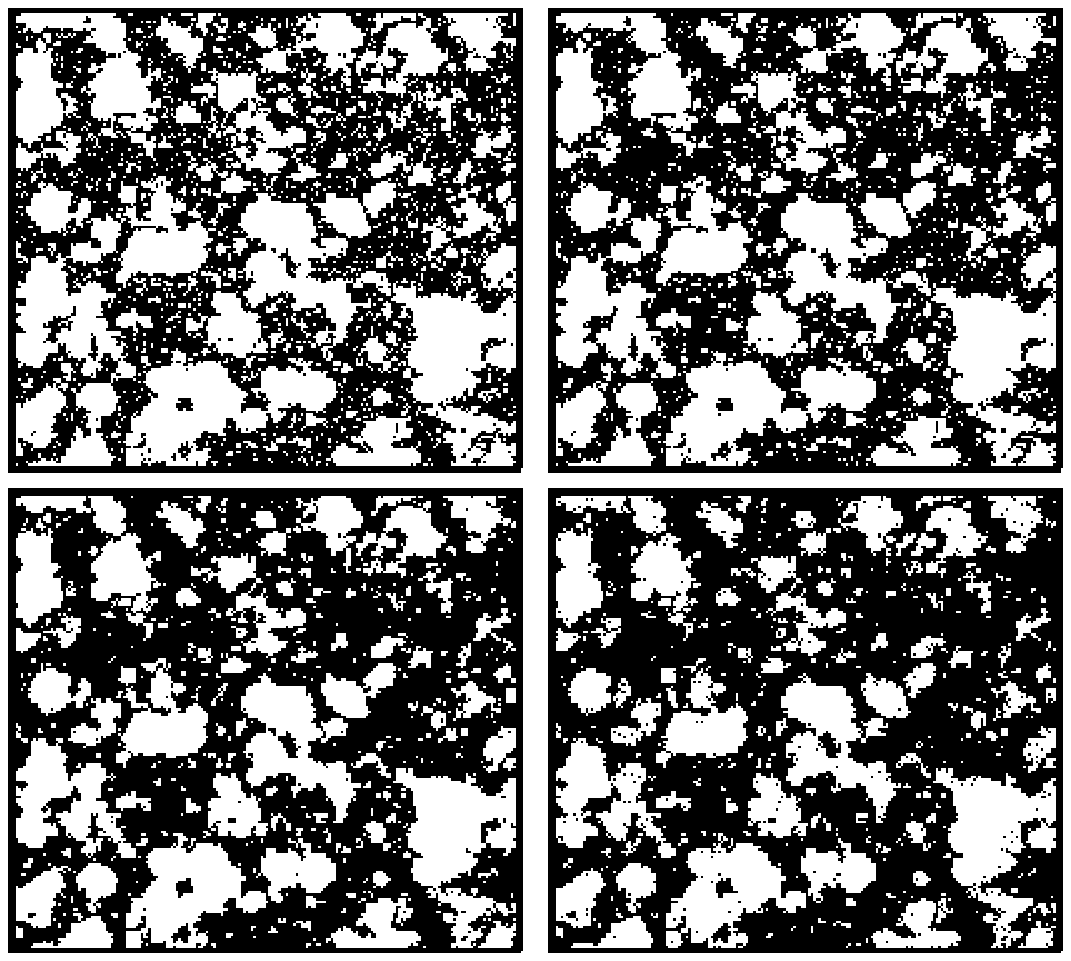}
\caption{\label{fig:threshold_depend} Distributions of optical depth at redshift $z = 9.3$, when
the average optical depth $\tau_0 = 4.4\times10^{-3}$. Regions with
relative optical depth $(\tau_0-\tau_\nu)/\tau_0 >$ threshold
(optically thin) are represented by white while other regions are
shown as black. The relative thresholds are 0.1, 0.3, 0.7, and 0.9
from left to right and top to bottom, respectively. These are plots
of slices in the simulation box with each side of 100 Mpc/h. }
\end{center}

\end{figure}

The uniformly distributed IGM would produce only uniform absorption
on the radio spectrum, it is the variation of density, ionization
state, and spin temperature which produces varied absorption that is
observable. This is illustrated in Fig.~\ref{fig:example}.
One may define an {\it absorber} or a {\it leaker} as a continuous
part of the spectrum for which
\begin{equation}
|\tau_\nu-\tau_0|>\tau_{th}
\end{equation}
where $\tau_0$ is the average optical depth over the whole
simulation box at that redshift, $\tau_{th}$ is the threshold value.
The strength of 21 cm absorption is typically not very strong due to
its weak transition coefficient. We take $\tau_{th}=0.002$, which
could induce features observable with high signal to noise
observation. The criteria $\tau_\nu-\tau_0 >\tau_{th}$ picks up
absolute absorbers while the criteria $\tau_0-\tau_\nu > \tau_{th}$
picks up absolute leakers.
Alternatively, one can also use a relative definition
\begin{equation}
\frac{|\tau_\nu-\tau_0|}{\tau_0}> R
\end{equation}
where $R$ is the threshold, and is taken as 0.5 here. We will see
later (in Fig.~\ref{fig:threshold_depend}) that the morphology and
distribution of the optically thin or thick regions remain nearly
the same if we use a different threshold. In this way, one can
select {\it relative absorbers} and {\it leakers}.

To characterize the strength of signals, we consider the equivalent
width (EW) of both absorbers or leakers. For absolute or relative
absorbers,
\begin{eqnarray}
W_\nu &=& \int_{absorber} \frac{|f_{\nu}-f_{IGM}|}{f_{IGM}} \, d\nu, \nonumber \\
&=&\int_{absorber} (1-e^{\tau_0-\tau_\nu}) \, d\nu
\end{eqnarray}
where $f_{IGM}$ is the residual flux of a high redshift radio source
after the mean absorption by the IGM, and $f_\nu$ is the flux at
frequency $\nu$. For absolute or relative leakers,
\begin{eqnarray}
W_\nu &=& \int_{leaker} (e^{\tau_0-\tau_\nu}-1) \, d\nu.
\end{eqnarray}
Clearly ionized regions will produce leakers and neutral clumps will
produce absorbers on high redshift quasar spectra.

Before using the equivalent width as our statistics, we test whether
it is a good tracer of the ionized (neutral) regions, and to what
extent we can extract $x_i$, $\rho$, and $T_{IGM}$ information from
EW statistics. This test includes two parts: (1) the consistency
between ionized regions and optically thin regions, and (2) the
correspondence between 3D bubbles and the projected 1D absorbers.

\begin{figure}[t]
\begin{center}
\includegraphics[width=\fwidth]{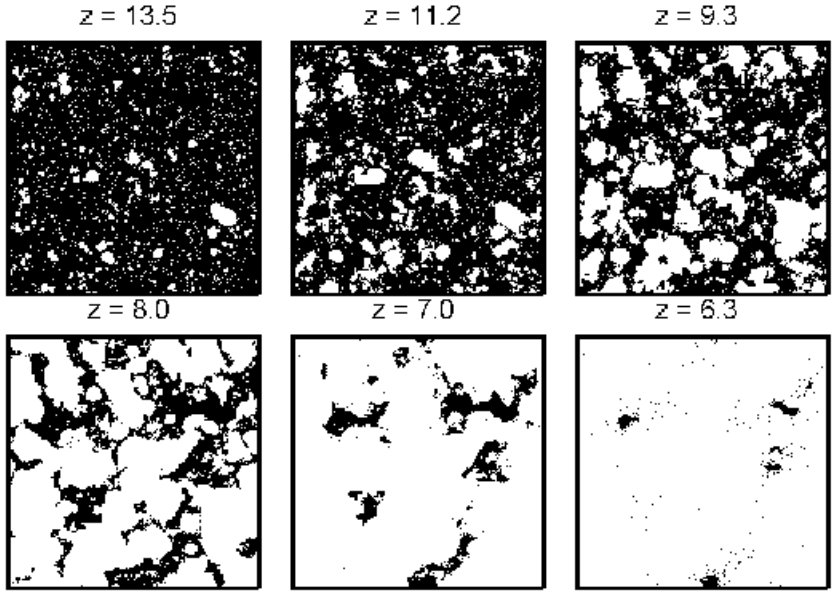}
\caption{\label{fig:tauslice_rel}Distributions of relative optical depth at redshifts $z
\sim$ 13.5, 11.2, 9.3, 8.0, 7.0, and 6.3. Regions with
$(\tau_0-\tau_\nu)/\tau_0 > 0.5$ (optically thin) are represented by
white while other regions are shown as black. These are plots of
slices in the simulation box with each side of 100 Mpc/h. The global
ionization fractions are same as Fig.~\ref{fig:xislice}. The
corresponding average optical depth for these redshifts are $\tau_0
= 1.1\times10^{-2}$, $7.5\times10^{-3}$, $4.4\times10^{-3}$,
$1.8\times10^{-3}$, $4.8\times10^{-4}$, and $6.7\times10^{-5}$
respectively.}
\end{center}

\end{figure}

\begin{figure}[h]
\begin{center}
\includegraphics[width=\fwidth]{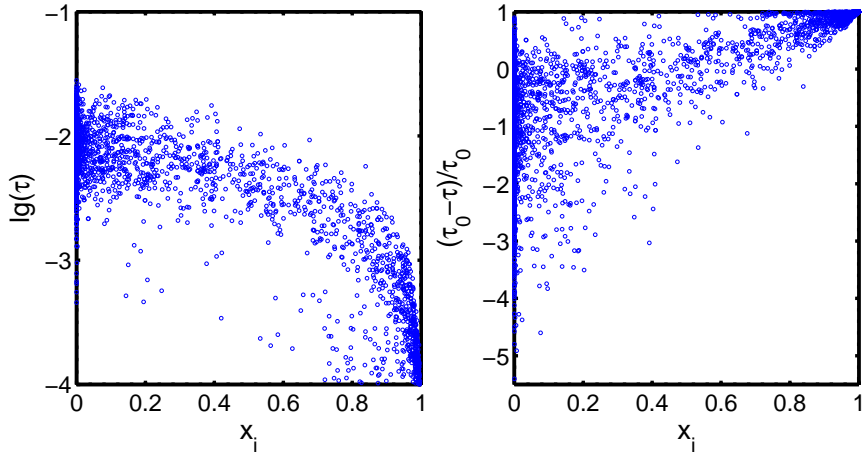}
\caption{\label{fig:scatterTauXi} Scatter plots of 21 cm optical
depth $\tau$ versus ionized fraction $x_i$ ({\it left panel}) and
the relative optical depth $(\tau_0-\tau)/\tau_0$ versus ionized
fraction $x_i$ ({\it right panel}). The redshift shown here is z =
9.3.}
\end{center}

\end{figure}

To do the first test, we display slices of optical depth field in
Fig.~\ref{fig:tauslice_rel}. In the optical depth calculation, the
spin temperature $T_s$ is taken to be the temperature output from
the simulation data. We have used the relative definition
$(\tau_0-\tau)/\tau_0 > 0.5$ to define the optically thin cells.
Optically thin regions are shown as white while optically thick
regions are shown as black. We see good agreement between optically
thin regions in Fig.~\ref{fig:tauslice_rel} and those highly ionized
regions in Fig.~\ref{fig:xislice}, and the configurations match very
well between the $x_i$-slices and $\tau$-slices. For more quantitative
comparison, we also made scatter plots of the absolute optical depth
$\tau$ and the relative optical depth $(\tau_0-\tau)/\tau_0$ versus
ionized fraction $x_i$ on a cell-by-cell basis in
Fig.\ref{fig:scatterTauXi} (for $z=9.3$). There is a strong correlation 
between the relative optical depth and the 
ionized fraction, as well as a correlation between the
 absolute optical depth and the ionization 
fraction at $x_i>0.6$. Note
that cells with high optical depth (large $\lg(\tau)$ or very negative
$(\tau_0-\tau)/\tau_0$) have low ionization fraction $x_i$, while high
threshold on $(\tau_0-\tau)/\tau_0$ picks out mostly highly
ionized regions.

\begin{figure}[htb]
\begin{center}
\includegraphics[width=\fwidth]{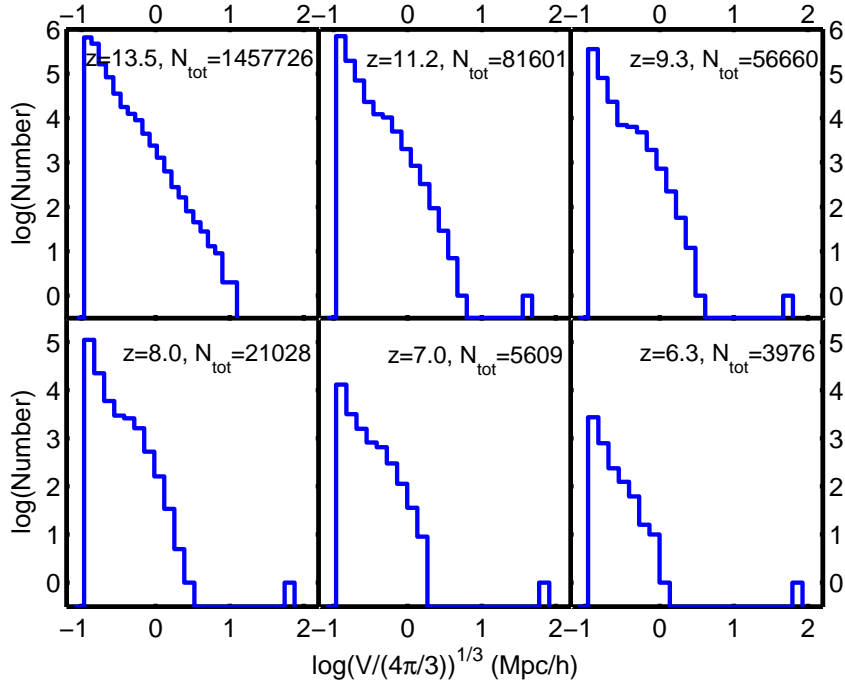}
\caption{\label{fig:tauBdist}Number distributions of relative
$\tau$-bubble volumes at redshifts $z =$ 13.5, 11.2, 9.3, 8.0, 7.0,
and 6.3, respectively. The criteria to select these bubbles is
$(\tau_0-\tau_\nu)/\tau_0 > 0.5$.}
\end{center}

\end{figure}

We may also use the FoF algorithm to select and connect cells by the
value of the optical depth instead of the ionization fraction. Size
distributions of the optically thin bubbles (denoted as relative
$\tau$-bubbles) are shown in Fig.~\ref{fig:tauBdist} for several
redshifts. The size distribution of these relative $\tau$-bubbles
are apparently very similar to that of the $x_i$-bubbles, but the
peaks are more smoothed, probably due to the mixture of ionization
state and density information in optical depth. So the optical depth
in the relative definition is a reasonable representative of the
ionization status of the IGM. Although it smears out the
characteristic scales of ionized bubbles, the $\tau$-field does
follow the large scale structure of the ionization field.

We also tried different relative thresholds in selecting these
 $\tau$-bubbles, in order to see how the signals would change with
 the thresholds we used. Fig.~\ref{fig:threshold_depend} shows the optical depth slices
at redshift 9.3 for thresholds $(\tau_0-\tau)/\tau_0 >$ 0.1, 0.3,
0.7, and 0.9, respectively. The shapes and distributions of the
bubbles are quite similar among these slices with different relative
optical depth thresholds. The $\tau$-slices with more strict
thresholds have more dark dots (optically thicker regions) inside
bubbles, and have less white dots (optically thinner regions)
outside bubbles. All these $\tau$-bubbles match well with
$x_i$-bubbles, and similar results can be found for other redshifts.
So the observational features in 21 cm forests would not change
significantly if different threshold is adopted.

\begin{figure}[htb]
\begin{center}
\includegraphics[width=\fwidth]{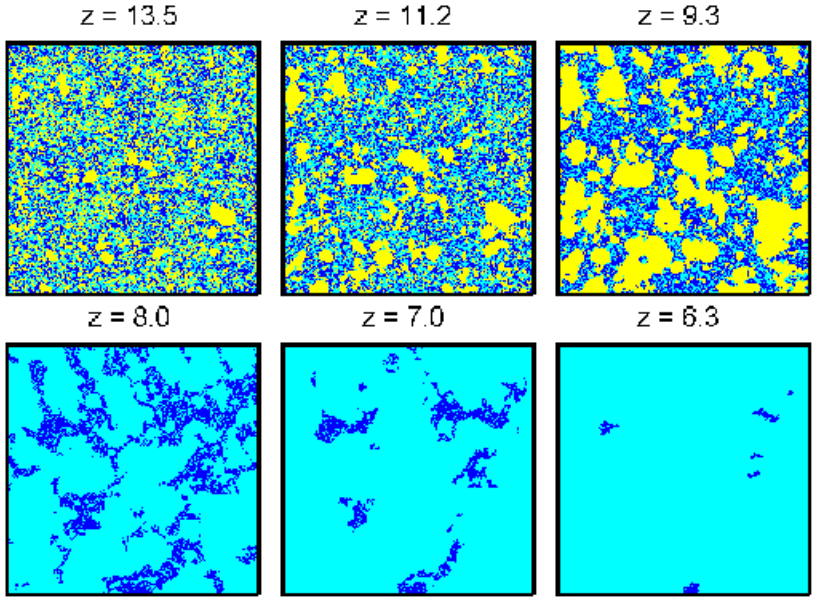}
\caption{\label{fig:tauslice_abs}
Distributions of absolute optical depth at six redshifts: $z =
13.5$, $z = 11.2$, $z = 9.3$, $z = 8.0$, $z = 7.0$, and $z = 6.3$.
Regions with $\tau_0-\tau_\nu > 0.002$ (optically thin) are
represented by yellow, regions with $\tau_\nu-\tau_0 > 0.002$
(optically thick) are shown as dark blue, and the other regions lie
in between are shown as light blue. The corresponding average
optical depth for these redshifts are $\tau_0 = 1.1\times10^{-2}$,
$7.5\times10^{-3}$, $4.4\times10^{-3}$, $1.8\times10^{-3}$,
$4.8\times10^{-4}$, and $6.7\times10^{-5}$ respectively. These are
plots of slices in the simulation box with each side of 100 Mpc/h. }
\end{center}

\end{figure}

\begin{figure}[htb]
\begin{center}
\includegraphics[width=\fwidth]{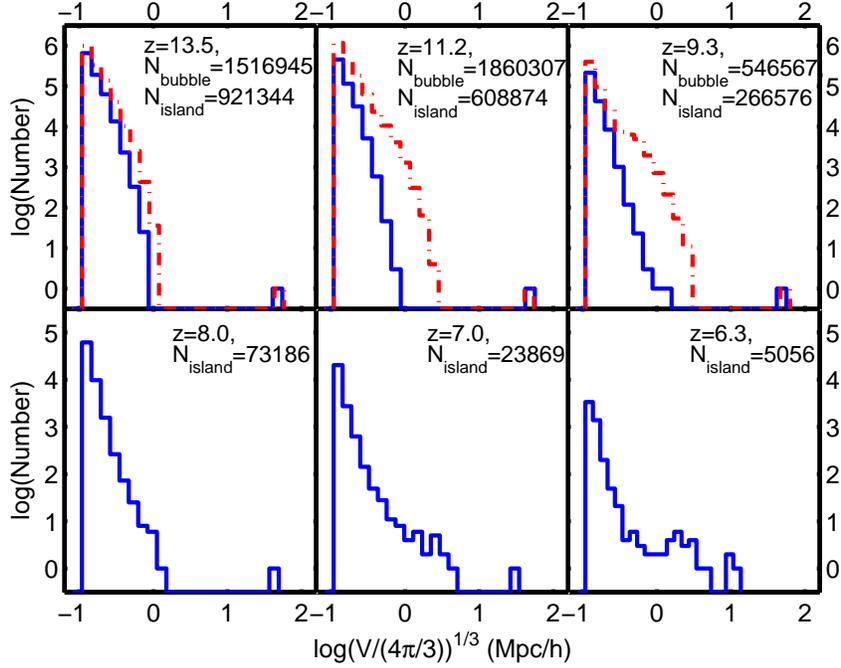}
\caption{\label{fig:tauBandIdist_abs}Size
distributions of absolute $\tau$-islands at redshifts $z =$ 13.5,
11.2, 9.3, 8.0, 7.0, and 6.3, respectively (blue solid lines), and
those of absolute $\tau$-bubbles at the three higher redshifts $z =$
13.5, 11.2, and 9.3 (red dot-dashed lines). The criteria to select
these islands is $\tau_\nu-\tau_0 > 0.002$, and the criteria for
these bubbles is $\tau_0-\tau_\nu > 0.002$.}
\end{center}
\end{figure}

\begin{figure}[htb]
\begin{center}
\includegraphics[width=\fwidth]{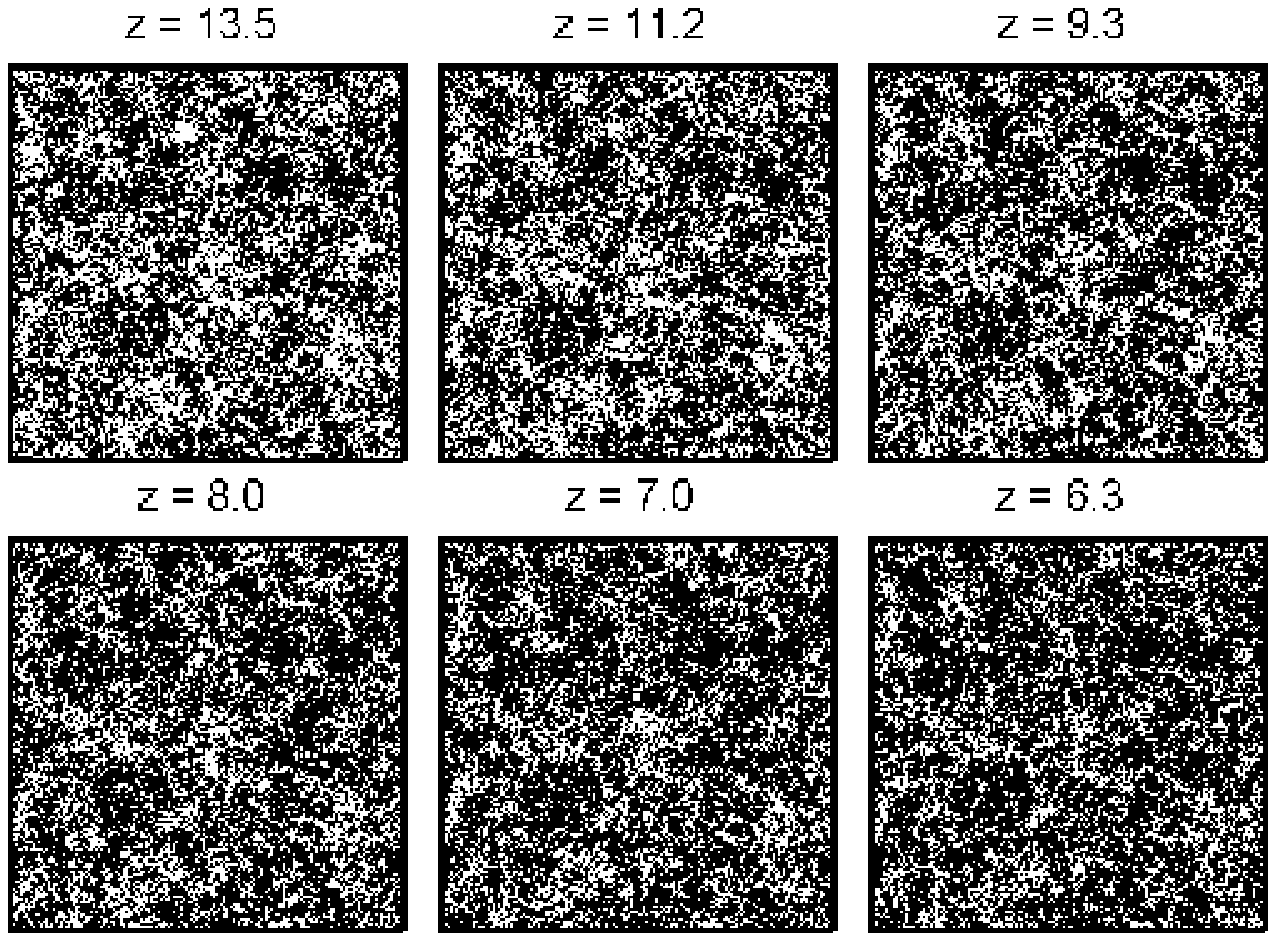}
\caption{\label{fig:den}Distributions of overdensity at different redshifts.
Overdense regions are represented by white while underdense regions
are shown as black.}
\end{center}
\end{figure}

Features picked up with absolute threshold are more relevant to real
observations. HII Bubbles and HI islands selected by the absolute
threshold $|\tau_\nu-\tau_0| > 0.002$ for the corresponding 6
redshifts are shown in Fig.~\ref{fig:tauslice_abs}. Using this
absolute optical depth threshold, there are both bubbles and islands
at three higher redshifts (i.e. 13.5, 11.2, and 9.3), which generate
leakers and absorbers on spectra respectively. At lower redshifts
there are only islands, since the average optical depth itself is
smaller than 0.002 (i.e. at redshifts 8.0, 7.0, and 6.3 here), so we
could only observe absorbers. For the leakers, smaller threshold
must be used in order to find them at these redshifts. The shapes
and distributions of these $\tau$-bubbles are similar to those
selected by relative thresholds. Combined with
Fig.~\ref{fig:xislice} and Fig.~\ref{fig:tauslice_rel}, we come to
the conclusion that no matter relative or absolute threshold we use,
the optical thin regions follow the highly ionized region very well,
at least on large scales. Size distributions of absolute
$\tau$-bubbles and $\tau$-islands for the three higher redshifts and
those of absolute $\tau$-islands for the other three lower redshifts
are shown in Fig.~\ref{fig:tauBandIdist_abs}.

In order to see whether the optical depth (hence the EW statistics)
is a good tracer of density fluctuations, we plot density slices in
Fig.~\ref{fig:den}. But because of the limited resolution of the
simulation (the overdensity is smoothed within each cell), the
density contrast is very small (typically $\delta < 1$) on these
scales, and do not show much evolution, and there is little
correlation with the ionization fraction. This means that we would
miss strong absorbers produced by small but highly dense regions
such as cosmic webs, this would affect the accuracy of this
simulation in representing the real case.

\begin{figure}[th]
\begin{center}
\includegraphics[width=\fwidth]{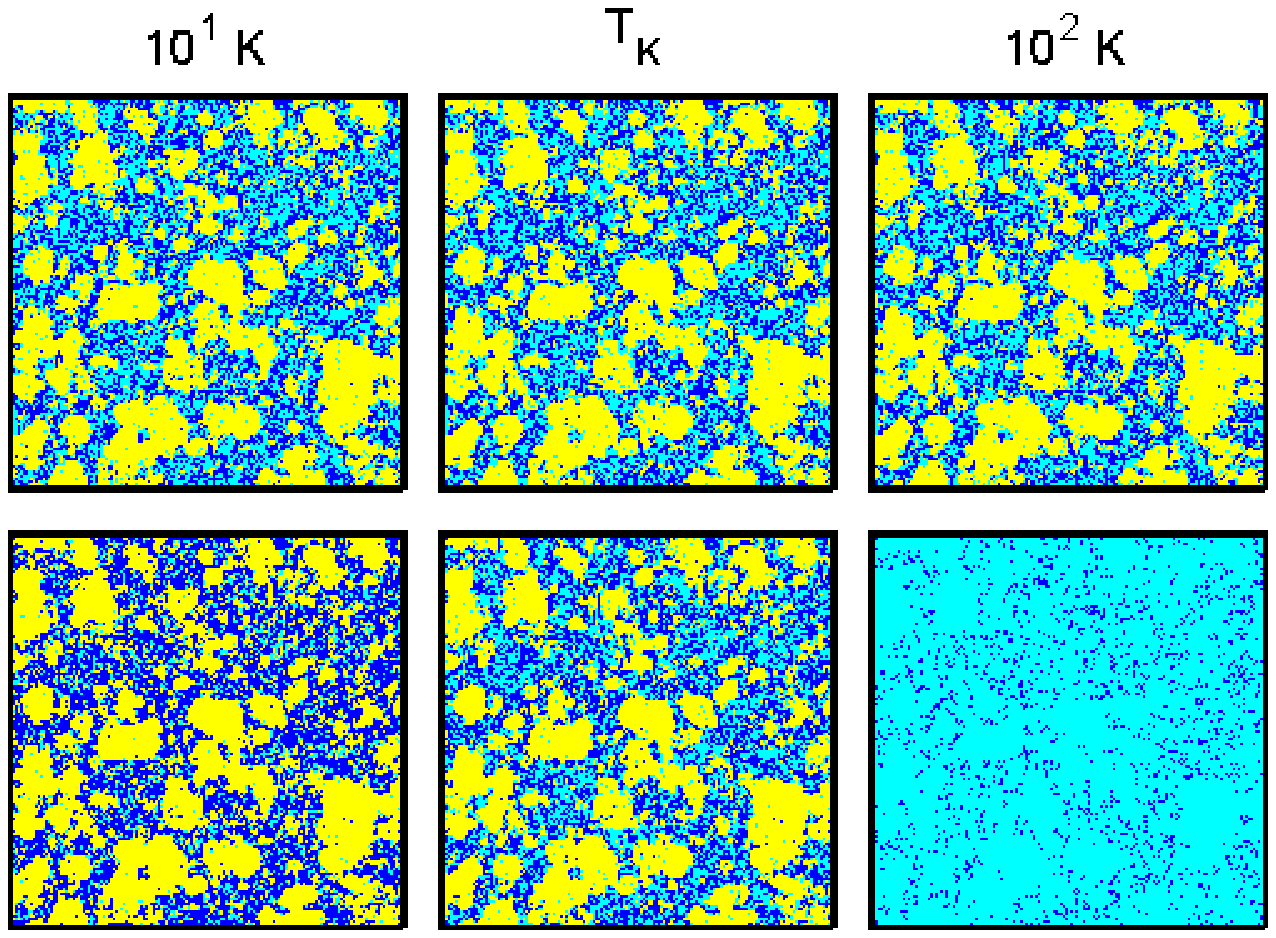}
\caption{\label{fig:temp} Distributions of optical depth with
different gas temperatures at redshift $z = 9.3$. Top panels:
optically thin regions with {\it relative threshold}
$(\tau_0-\tau_\nu)/\tau_0 \ge 0.5$ are shown as yellow, optically
thick regions with $(\tau_\nu-\tau_0)/\tau_0 \ge 0.5$ are shown as
dark blue, while other regions in between are labeled as light blue.
The gas temperatures are assumed to be $10^1$K, gas temperature
based on velocity dispersion, and $10^2$K, from left to right,
respectively. Bottom panels: regions with {\it absolute threshold}
$\tau_0-\tau_\nu > 0.002$ are shown as yellow, regions with
$\tau_\nu-\tau_0 > 0.002$ are shown as dark blue, while other
regions are shown as light blue. The temperatures are the same as
top panels. }
\end{center}
\end{figure}

Can we get some information about the temperature of the IGM from 21
cm forest signals? Although there is no heating information
incorporated in the simulation, and the gas temperature is derived
from the velocity dispersion of the particles, it is still possible
to test the sensitivity of optical depth and the forest signal to
the IGM temperature, by assuming several different gas temperatures.
In top panels of Fig.~\ref{fig:temp}, we plot relative $\tau$-slices
at redshift 9.3 for the IGM temperatures of $10^1$K, the gas
temperature from simulation, and $10^2$K, from left to right,
respectively (in this excise we set the same temperature at every point
within the simulation volume). Absolute $\tau$-slices with
the corresponding gas temperatures are plotted in bottom panels of
Fig.~\ref{fig:temp}.

Obviously, $\tau$-slices with relative thresholds are insensitive to
the gas temperature, while the absolute $\tau$-slices are very
sensitive. When we raise the IGM temperature to be $10^2$K, the
average optical depth at redshift 9.3 falls below the threshold
0.002, and no optical depth could drop below the mean value by
0.002, so there is no optically thin region in the absolute
definition in this redshift, just as the three slices of lower
redshifts in Fig.~\ref{fig:tauslice_abs}. If we raise the IGM
temperature further to be $10^3$K, then very few regions could have
optical depth larger than the average depth by 0.002. Hence, we
could easily extract $T_{K}$ information from absolute $\tau$
threshold observations while ionization state of the IGM could be
extracted from relative $\tau$ threshold observations.

It is very interesting to find a method to measure the IGM
temperature during the epoch of reionization. While the kinetic
temperature of the IGM and spin temperature of hydrogen are usually
assumed to be much higher than the CMB temperature in 21 cm power
spectrum analysis, and the $T_{K}$ information of the IGM is totally
erased in ``21 cm tomography'' observations, the ``21 cm forest''
observations serve as an excellent tool to separate the $T_{K}$
information from the density and ionization status.

\begin{figure}[t]
\begin{center}
\includegraphics[width=\fwidth]{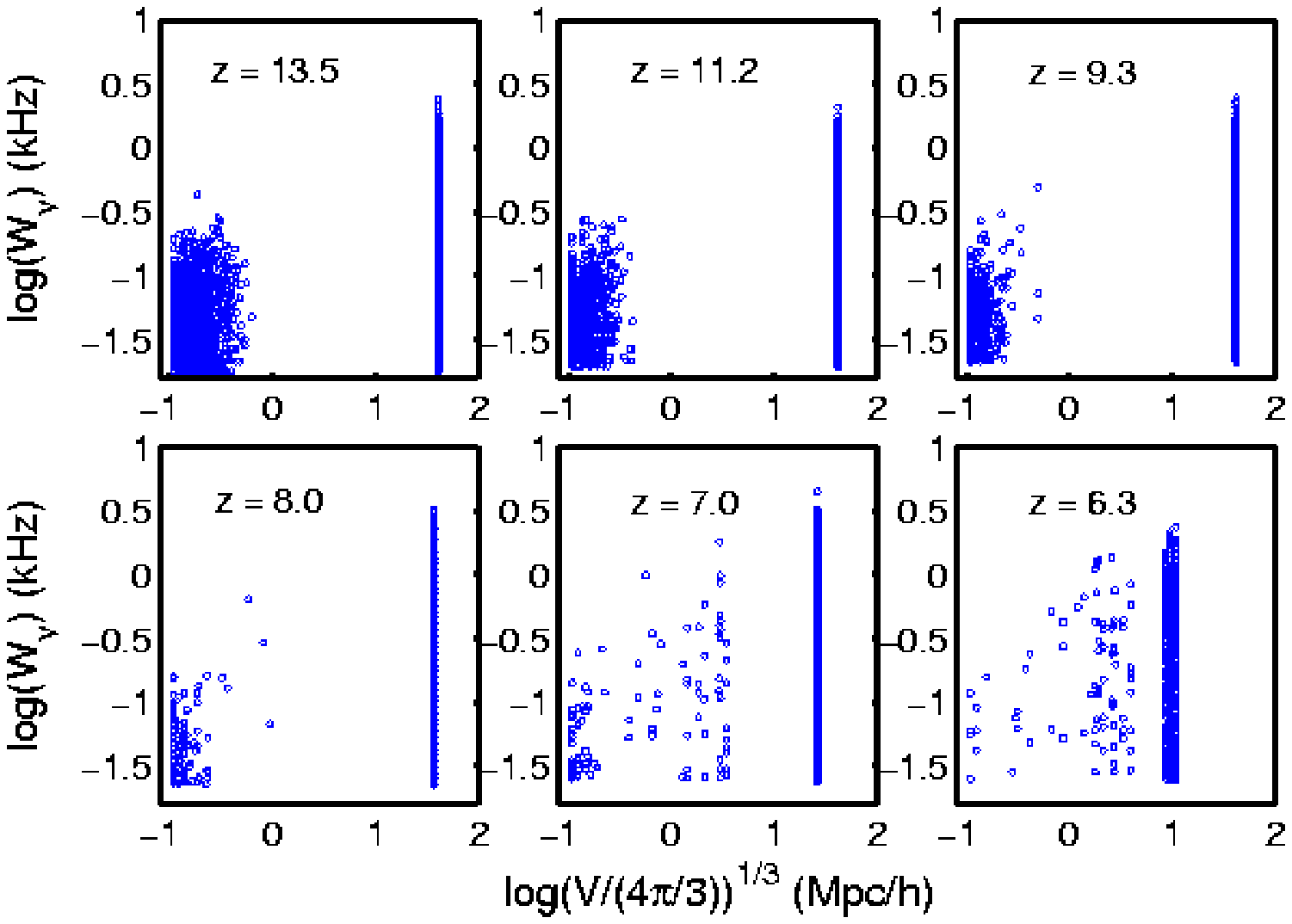}
\caption{\label{fig:abs_absorber_scatter}Scatter
plots of volumes of $\tau$-islands which hold absorbers versus EW of
those absorbers at different reshifts.}
\end{center}

\end{figure}

Now we have seen some correspondence between the optical depth and
the ionized fraction selected bubbles. However, in order to
guarantee the validity of our EW statistics, we should also test the
correspondence between observed absorbers and $\tau$-islands, or
leakers and $\tau$-bubbles. To do so, we put random lines of sight
through the simulation box, calculated the EW of each absolute
absorber, and recorded their center positions. Then we checked
whether the centers of absorbers reside in those $\tau$-islands we
found. Results show that most EW features have their centers within
$\tau$-islands, and those EW features which are not in
$\tau$-islands are in one-cell-islands, which were not defined as
islands in our ``FOF'' algorithm. Fig.~\ref{fig:abs_absorber_scatter}
shows scatter plots of absorbers' equivalent width and the volumes
of their corresponding $\tau$-islands in which the centers of these
absorb features reside. Absolute definition is used here for
absorbers and islands, as is relevant in real observations. Although
there is some correlation between the volume and equivalent width,
we see that the scatter is very large, as the line of sight may pass
through the region far from the center. One can also see a spike in
the volume distribution, which corresponds to the connected HI
region (partially ionized at lower redshifts) in the simulation box.
This connected island does not disappear until the overall optical
depth drops as seen at $z = 7.0$ when some places in the island
failed to connect those regions around them and give birth to many
small islands. More clearly, at $z = 6.3$, one connected island
fragmented into three big islands and many smaller ones. They exist
till late stages of reionization because of the self-shielding
effect of these dense regions.

In summary, optical depth can be a reasonable tracer of the ionized
fraction and the equivalent width could serves as a great probe of
cosmic reionization. While the features picked with relative
thresholds are better in extracting ionization information, the
features picked by absolute thresholds are very sensitive to the IGM
temperature. However, the small scale density fluctuation is not
incorporated in our simulation, and the density field on large
scales plays a minor role in determining the optical depth. So we do
not expect to get information about local density fluctuation from
EW statistics. In addition, we see correspondence between the
3-dimensional bubbles and the projected 21 cm forest signals.

\section{Statistical Distribution of 21 cm Forest Signals}

\begin{figure}[h]
\begin{center}
\includegraphics[width=\fwidth]{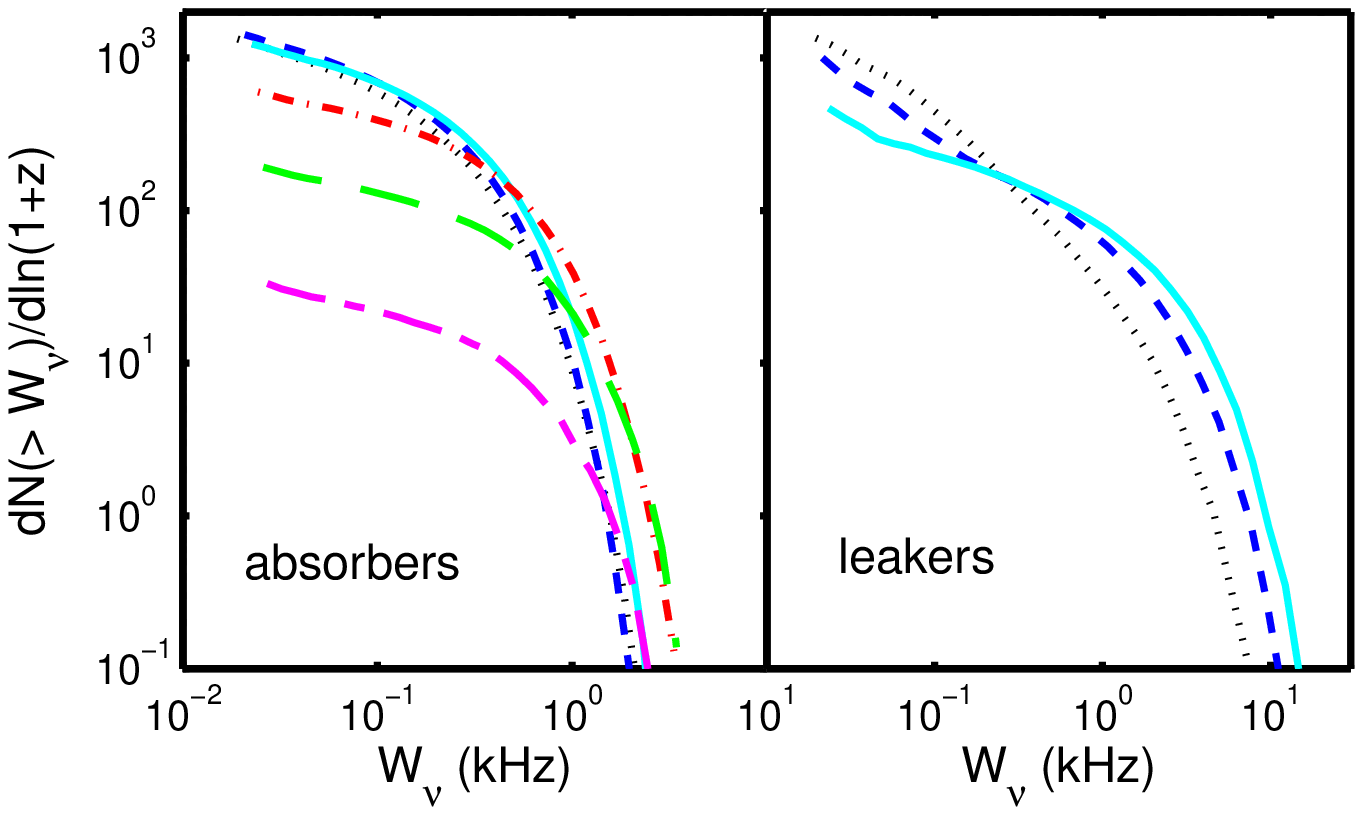}
\caption{\label{fig:Num}The number density of absolute absorbers (left panel) and
leakers (right panel) with equivalent width larger than $W_\nu$, as
a function of $W_\nu$. The curves for redshifts 13.5, 11.2, 9.3,
8.0, 7.0, and 6.3 are represented by dotted (black) line,
short-dashed (blue) line, solid (cyan) line, dot-dashed (red) line,
long-dashed line (green), and long-and-short-dashed (purple) line
respectively for absorbers, and only the three curves for higher
redshifts are shown for leakers with the same line types as those
for absorbers at the same redshifts.}
\end{center}
\end{figure}

First we calculate the average optical depth $\tau_0$ over the whole
simulation box for each redshift $z$, and take this $\tau_0$ as the
optical depth of the intergalactic medium, $\tau_{IGM}$. Then we put
400 random lines of sights (LOS) through each side of the simulation
box at a specific redshift $z$, and calculate the optical depths
along each line of sight as a function of observed frequency.
Using the threshold $(\tau_\nu-\tau_0) > 0.002$, we pick every
absorber along each line of sight, then calculate and record the EWs
of them for every random LOS. The periodic boundary condition is
employed here.

The number densities of absorbers and leakers are shown in
Fig.~\ref{fig:Num}. The total number of these absorbers decreases
significantly with decreasing redshift throughout the reionization.
This is especially obvious for absorbers of small equivalent width,
since the small islands were being ionized and becoming parts of
ionized bubbles as the reionization proceeded. However, the number
of absorbers with the largest equivalent width does not change as
much, and the largest absorbers seem to persist to lowest redshifts.
This is due to the self-shielding effect of the dense neutral
regions, which precisely generated the spikes seen in
Fig.~\ref{fig:abs_absorber_scatter}. They give us an opportunity to
observe large absorbers even at lower redshifts. For the leakers,
the number of smaller ones decreases with decreasing redshift since
those small ionized bubbles were merging to generate larger ones,
while the number density of larger leakers increases.

\begin{figure}
\begin{center}
\includegraphics[width=\fwidth]{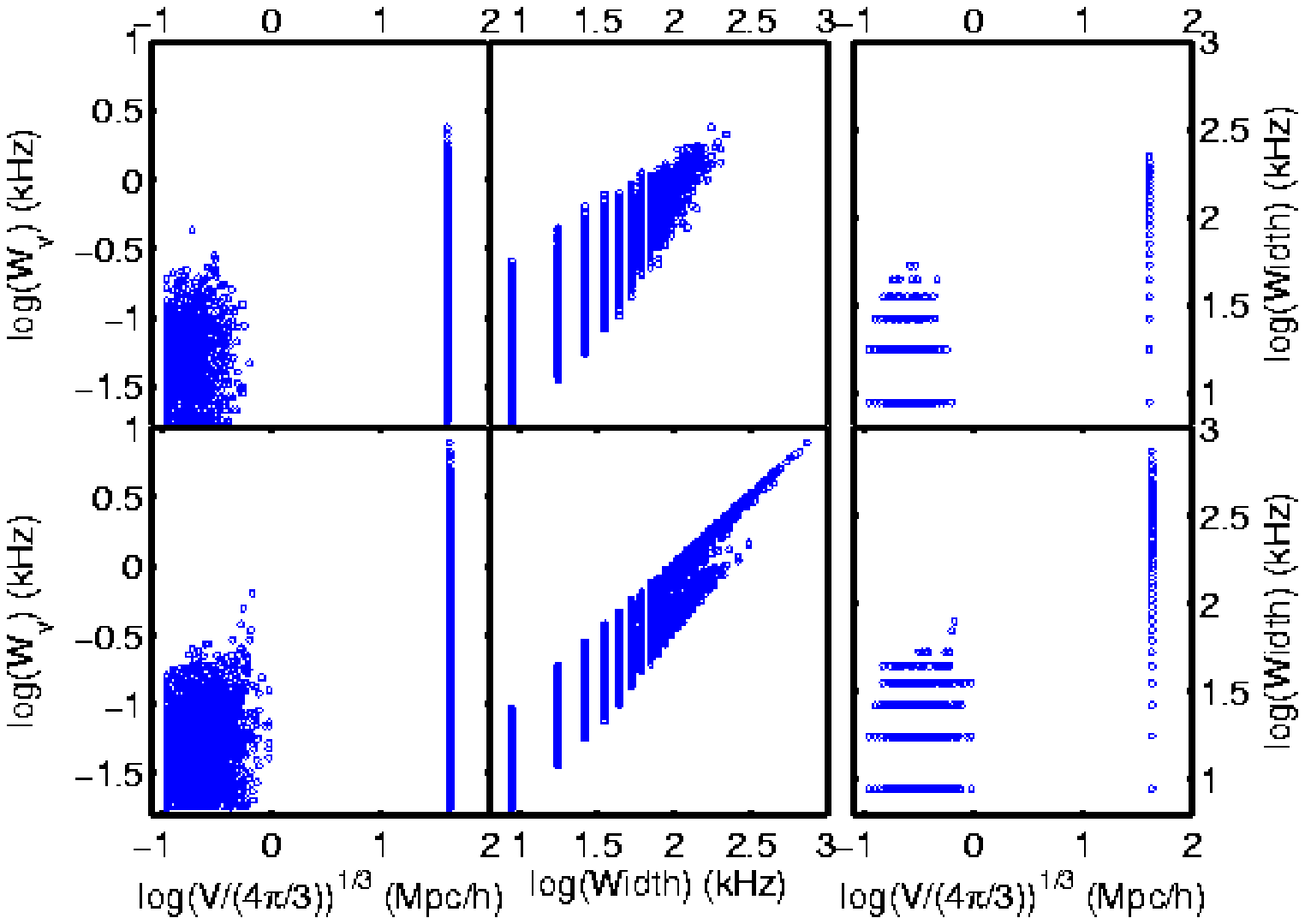}
\caption{\label{fig:3Dscatter} Top panels:
the correspondence among the absorbers' width, their equivalent
width, and the volumes of $\tau$-islands which created them. Bottom
panels: the correspondence among the leakers' width, their
equivalent width, and the volumes of $\tau$-bubbles which created
them. The redshift shown here is 13.5. }
\end{center}
\end{figure}

Besides the equivalent width, the width of these absorbers or
leakers is also an observable in radio probes, and it has good
correspondence to the equivalent width as shown in
Fig.~\ref{fig:3Dscatter}. Just as EWs of absorbers and their
corresponding $\tau$-islands were shown in
Fig.~\ref{fig:abs_absorber_scatter}, Fig.~\ref{fig:3Dscatter} shows
the consistency between feature widths and the $\tau$-islands or
$\tau$-bubbles associated with them. The maximum width of an
absorber created by each island has a stronger correlation with the
island size, and the same applies to relation between bubble size
and the maximum width of leakers.

\begin{figure}[t]
\begin{center}
\includegraphics[width=\fwidth]{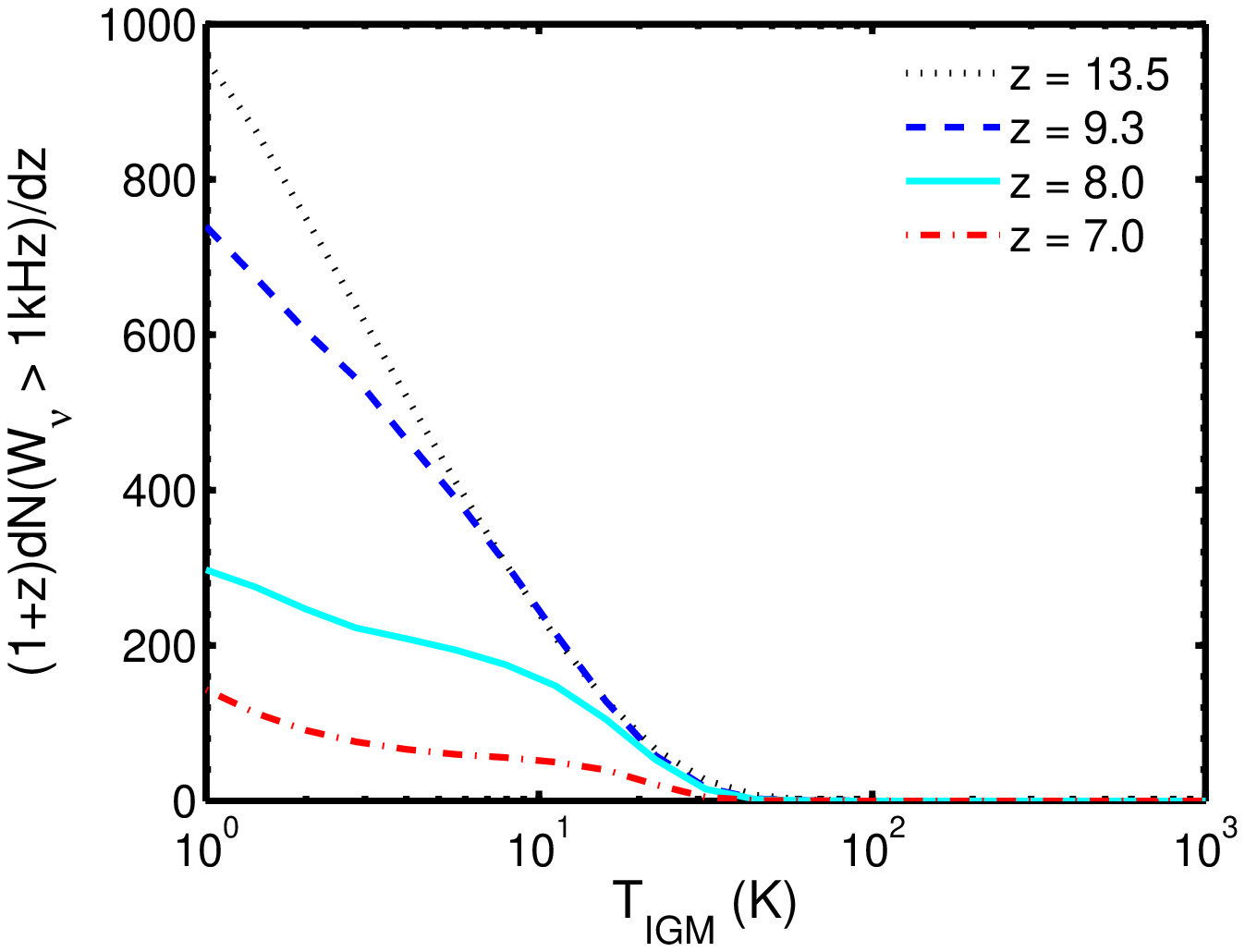}
\caption{\label{fig:NonTemp}Number density of absolute absorbers with equivalent width
$W_\nu > 1\kHz$ as a function of the IGM temperature. The redshifts
are $z = 13.5 \,(dotted, black)$, $z = 9.3 \,(dashed, blue)$, $z =
8.0 \,(solid, cyan)$, and $z = 7.0 \,(dot-dashed, red)$,
respectively.}
\end{center}
\end{figure}

The optical depth is very sensitive to the IGM temperature. We
calculated the number density of absolute leakers that have
$W_\nu> 1\kHz$ as a function of the gas temperature. The results for
several redshifts are presented in Fig.~\ref{fig:NonTemp}. The
number densities of potentially observable absorbers decrease
rapidly as the temperature increases. This means, on one hand, with
the first generation of low frequency equipments, we may not be able
to observe the signals that we are looking for if the IGM was really
heated up, but on the other hand, we could obtain information on the
gas temperature at the epoch of reionization from 21 cm forest
observations. This is important because the IGM temperature contains
very rich information about early structure formation, including the
initial mass function of the first stars, properties of early
mini-quasars and related X-ray heating, etc. In this sense, the 21
cm forest can be a powerful tool not only to probe the IGM in the
early universe, but also to give constraints on the properties of
early structures.

Our prediction of the mean IGM optical depth is a little smaller
than those by \citet{Carilli02} at higher redshifts but larger than
theirs at lower redshifts. For example, at $z = 12$
\citet{Carilli02} found that $\tau_{IGM} \sim 1\%$, and some of the
high-density regions reach $\tau \sim 5\%$ and above; by $z \sim 8$,
the mean IGM optical depth has dropped to $\tau_{IGM} \sim 0.1\%$.
In our simulation, however, $\tau_{IGM} \sim 0.9\%$ at $z \sim 12$
and it only drops to $\tau_{IGM} \sim 0.18\%$ at $z = 8$.

However, unlike \citet{Carilli02}, we find very few strong
absorbers. In our simulation box, only about one in two
million cells would have $\tau
\ge 5\%$ at $z \sim 12$, and no absorbers has $\tau \ge 2\%$ at $z
\le 10$, while \citet{Carilli02} still predicted 50 lines per unit
redshift with $\tau \ge 2\%$ at $z = 10$ and about 4 lines per unit
redshift at $z = 8$. Some of these differences may be due to the different
values of cosmological parameters adopted in the two simulation. However,
the main difference probably comes from the different
box size of the simulations. \citet{Carilli02} used a simulation
with smaller volume but higher spatial resolution \citep{Gnedin00},
in which the typical values of the cosmic overdensity $\delta \sim
10$. Our simulation has a much larger box volume, which is much more
reliable for large scale reionization, but the size of each cell is
larger. This affects the density fluctuation dramatically, and our
typical overdensity within each cell is $\delta < 1$, with a few
regions have $\delta > 1$. As seen in Fig.~\ref{fig:den}, the
density field evolves very little when smoothed within the cells,
though at smaller scales there must be some evolution in density
contrast. Therefore, the resultant local optical depth is
underestimated.

On the other hand, the first HII regions start to appear at redshift
about 8 in the simulation by \citet{Gnedin00}, which is much later
than that in our simulation \citep{STC08}. As a result,
\citet{Carilli02} found higher optical depth of 21 cm forest than us
at high redshift. Further, in \citet{Carilli02}, those absorbers are
typically filaments. They are at first shock-heated to about 100 K,
and at $z < 10$ Ly$\alpha$ heating increases the mean spin
temperature to above 100 K. In our simulation, the resolution is not
high enough to capture the filaments anyway. The Ly$\alpha$ heating
is also unlikely to heat the gas to so high a temperature, more
likely it could only heat up the IGM to a temperature below 100 K
before reionization \citep{CM04,Y09}. In the simulation, these
heating mechanisms are not incorporated, and the temperature is just
from the velocity dispersion of the particles. Therefore we have
lower spin temperature and this results in the higher mean IGM
optical depth at lower redshifts, though we can artificially adjust
the gas temperature as a proxy for incorporating the various heating
mechanisms.

\section{The number of high redshift radio sources}

To observe the absorption or leak features in the spectrum, the
spectrum of the point source has to be observed with a certain
precision, which depends on the brightness of the quasar. The
minimum detectable flux density of an interferometer is related to
the system temperature $T_{\rm sys}$, the effective aperture area
$A_{\rm eff}$, channel width $\Delta\nu_{\rm ch}$, integration time
$t$, and the signal-to-noise ratio $S/N$ by:
\begin{equation}
\Delta S_{\rm min}\, =\, \frac{2\, k_{\rm B}\, T_{\rm sys}}{A_{\rm
eff}\sqrt{\Delta\nu_{\rm ch}\,t}}\, \frac{S}{N}.
\end{equation}
For the observation of resolved signals (i.e. the resolution pixel
$\Delta\nu_{\rm ch}$ is narrower than the width of the feature to be
observed), the lower limit of the flux density of background sources
for an observation of absorbers or leakers with $|\tau_\nu-\tau_0|
\geq \tau_{th}$ (when the $\tau_\nu$, $\tau_0$, and $\tau_{th}$ are
all very small) is:
\begin{eqnarray}
S_{\rm min}&=&181.8 \mJy\left(\frac{S/N}{5}\right)\left(
\frac{0.002}{\tau_{\rm th}}\right) \left(\frac{1
\kHz}{\Delta\nu_{\rm ch}}\right)^{1/2}
\nonumber\\
&&\times \left(\frac{2\times10^3\,\m^2\K^{-1}}{A_{\rm eff}/T_{\rm
sys}}\right) \left(\frac{100\, \hr}{t}\right)^{1/2}.
\end{eqnarray}
where the ratio $A_{\rm eff}/T_{\rm sys}$ is an intrinsic parameter
describing the sensitivity of an interferometry array. The
systematic temperature is the sum of the receiver noise and the
temperature of the uncleaned foregrounds.

However, the optical depths of the ``absorbers'' and ``leakers'' are
small, while the signals are relatively wide($10^1\sim 10^2\kHz$),
so that observation of unresolved signals (or just below resolution)
is more promising. In this case, the lower limit of the flux density
for an observation of absorbers or leakers with $W_\nu \geq W_{th}$
is:
\begin{eqnarray}
S_{\rm min}&=&11.5 \mJy
\left(\frac{S/N}{5}\right)\left(\frac{1\kHz}{W_{\rm th}}\right)
\left(\frac{\Delta\nu_{\rm ch}}{1 \MHz}\right)^{1/2}
\nonumber\\
&&\times \left(\frac{2\times10^3\m^2\K^{-1}}{A_{\rm eff}/T_{\rm
sys}}\right) \left(\frac{100 \hr}{t}\right)^{1/2}.
\end{eqnarray}
The demanded flux density of the background source is much lower
compared to the resolved observations, making more high redshift
radio sources qualified for such observations.

We now investigate how many such quasars can be observed.
Specifically, we calculate the number of quasars with flux density
at the observed frequency $\nu = 1420.4/(1+z)\,\MHz$ larger than the
lower limit described above. In the following, we assume that all
the radio quasars are of the same spectral energy distribution with
that of  the powerful radio galaxy Cygnus A, i.e.
$P_{\nu}\propto\nu^{-1.05}$ \citep{Carilli02}. Then the limiting
flux density converts to a limiting luminosity density at a
rest-frame frequency of 151 MHz, $P_{151}^{\rm min}$.

\citet{Jarvis01} have pointed out that at $z=4$, the comoving number
density of radio sources with luminosity density at the frequency
151 MHz, $P_{151}\geq 6\times10^{35}\,\ergs \,\psec \Hz^{-1}$, is
$2.4\times 10^{-9}\,\Mpc^{-3}$. We extrapolate this number density
to higher redshift and lower luminosity. Assuming that the comoving
number density of qualified quasars at redshift $z$ is given by
$n_{\rm CM}(P_{151}\geq P_{151}^{\rm min},z)$, and according to the
best-fitting luminosity function given by \citet{Jarvis01}, we have
\begin{equation}
dn_{\rm CM}(P_{151},z)/dP_{151} \,=\, C(z)\, P_{151}^{-2.2}.
\end{equation}
where $C(z)$ is a normalization constant, which in general is a
function of redshift. But it is suggested by \citet{Jarvis01} that
there is very little evidence for an abrupt decline in the comoving
space density of the most luminous low-frequency-selected sources at
high redshift, and the best-fitting model gives a constant comoving
space density.
If, optimistically, we hypothesize
quasars evolve according to the flat model, i.e. the number density
does not vary with redshift in the comoving coordinate, then the
number of qualified quasars ($P_{151}\geq P_{151}^{\rm min}$) in the
whole sky per redshift interval is:
\begin{equation}
\frac{dN }{dz}\, =\, \frac{C}{1.2}\, {P_{151}^{\rm min}}^{-1.2}
\times 4 \pi r_{\rm CM}^2(z)\, \frac{dr_{\rm CM}(z)}{dz},
\end{equation}
where $r_{\rm CM}(z)=\int_0^z\,c\,dz/H(z)$ is the comoving
coordinate corresponding to redshift z. On the other hand, based on
optically selected quasars from SDSS, \citet{Cirasuolo06} suggest
that independent of the adopted radio spectral index, the drop in
the space density of optically bright radio loud quasars between $z
\sim 2$ and $z \sim 4.4$ is at most a factor $\sim 1.5 - 2$. If we
adopt the upmost factor 2, then
\begin{eqnarray}
C(z)&=& C(z=4)\, \exp[-\,\frac{\ln{2}}{2.4}\,(z-4)].
\end{eqnarray}

\begin{figure}[h]
\begin{center}
\includegraphics[width=\fwidth]{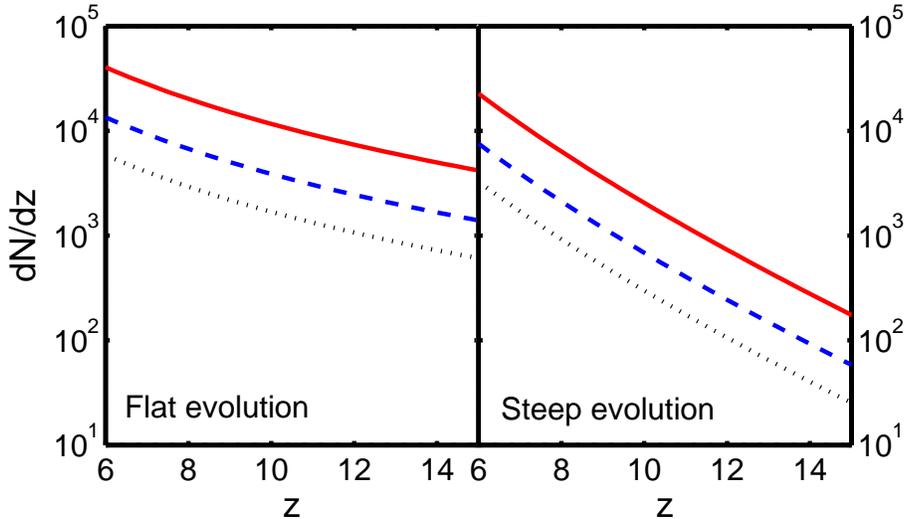}
\caption{\label{fig:qso_evo}The number of quasars in the whole sky that
can be used to
detect signals with $W_\nu \geq 1$kHz per redshift interval. {\it
Left panel}: the number of qualified quasars in the flat evolution
model. {\it Right panel}: the number in the steep evolution model.
The sensitivity of the radio array is taken to be $A_{\rm
eff}/T_{\rm sys} = 5000$ $\m^2\K^{-1}$, $2000$ $\m^2\K^{-1}$, and
$1000$ $\m^2\K^{-1}$ for the {\it red-solid}, {\it blue-dashed}, and
{\it black dotted} line, respectively. The integration time is
assumed to be 100 hours here.}
\end{center}
\end{figure}

We plot the number of quasars which satisfy the above criteria for
probing signals with  $W_{\nu}\geq 1$ kHz in Fig.~\ref{fig:qso_evo}
for the flat evolution case in the left panel, and for the evolved
case in the right panel. Here we choose three values of sensitivity,
$A_{\rm eff}/T_{\rm sys} = 5000$ $\m^2\K^{-1}$, $2000$
$\m^2\K^{-1}$, and $1000$ $\m^2\K^{-1}$ respectively, which can be
achieved in future observation programs. The Square Kilometer Array
(SKA) will have $A_{\rm eff}/T_{\rm sys} = 5000$ $\m^2\K^{-1}$ at
the frequency between 70 MHz and 300 MHz (see
http://www.skatelescope.org/), and the Low Frequency Array (LOFAR)
could marginally achieve $A_{\rm eff}/T_{\rm sys} = 1000$
$\m^2\K^{-1}$ at 200 MHz (http://www.lofar.org/). We can see that
for the flatly evolved quasar density, there would be many bright
quasars which could be used as our background sources. However, for
the steeply evolved quasar model, those suitable background quasars
would be very few.

\section{Conclusions}
We used a numerical simulation with star formation and radiative
transfer \citep{STC08} to study the 21 cm forest signals during the
epoch of reionization. We first defined absolute and relative
absorbers and leakers on spectra of background sources produced by
structures along the line of sight, and checked whether the
absorbers or leakers can represent the ionization and thermal state
of the IGM. We found that the optical depth can be a reasonable
tracer of the ionized fraction of the IGM, and the equivalent width
could serves as a great probe of cosmic reionization. The features
selected by absolute threshold in optical depth are also very
sensitive to the IGM temperature, so the temperature information of
the IGM could be separated from the ionization fraction and density
fluctuations with 21 cm forest observation, making it a good
complement to 21 cm tomography observation. At the scales we
studied,  the density fluctuation plays a minor role in determining
the optical depth. Also, we see correspondence between the
3-dimensional bubbles and the projected 21 cm forest signals.

The number densities of leakers and absorbers with different
equivalent widths evolve with redshift, showing the evolution of
ionization status of the IGM. While the total number of absorbers
decreases significantly with decreasing redshift, the largest
absorbers persist to low redshifts because of the self-shielding
effect. As for the leakers, diminished small ones and ballooning
large ones with time show the signature of merging ionized bubbles.
From the largest width of leakers or absorbers in 21 cm forest
spectra, we may infer the size of ionized bubbles or neutral islands
at specific redshifts.

The most important advantage regarding the 21 cm forest observation
is its sensitive dependence of signals on the IGM temperature. The
number density of potentially observable signals decreases
dramatically with the increasing gas temperature, so the temperature
of the IGM at each redshift could be constrained potentially through
the 21 cm forest observations. As the thermal history of the IGM
carries rich information about the first luminous sources, including
the population of the first stars, properties of mini-quasars, and
complex heating processes, the 21 cm forest observation is a
potentially powerful probe of the early structure formation of the
universe. The measurement of the gas temperature would also help
improve the precision of the measurement of cosmological parameters.

We also compared our results with a previous work by
\cite{Carilli02}, who used a simulation with higher resolution but
smaller volume \citep{Gnedin00}. While our simulation has a much
larger volume which is necessary to account for large HII regions at
late epochs of reionization, small-scale structures such as
small filaments and minihalos are not resolved, especially in the
binned data used for post-analysis. With higher resolution simulations, higher gas
overdensities and hence higher optical depths would be expected.
Since heating processes are not included in our simulation, the assumed gas temperature is not realistic, but we do consider a range of possible temperatures. With these limitations, we can not make an accurate prediction for the signal, but instead provide a basic expected range. Nonetheless, our simulation does reveal some qualitative features of the large scale 21cm signal, and propose here that the 21 cm forest observations, especially the EW statistics, could play an important role in extracting temperature information of the IGM. We reserve more realistic investigation of the 21 cm forest to future works, for example by using higher-resolution versions of the hydro+radiative transfer simulations presented in \citet{TCL08}, which have self-consistent hydrodynamics and temperature evolution.

Finally we discussed the the number of luminous high redshift radio sources
which is potentially available for observing such forest signals. We investigated the
requirement on the luminosity of background radio
sources for the observation of signals with equivalent widths larger
than 1kHz. We considered how the number of qualified quasars in the
whole sky evolves with redshift according to two models.
If quasars evolves according to the flat evolution model, we would
have sufficient quasars as background sources at high redshift; if
quasars evolve fast, however, only those signals at relatively lower
redshift can be observed in the near future.

\acknowledgements
We thank Bin Yue, Andrea Ferrara, Steven Furlanetto and Chris Carilli
for helpful discussion. This research is supported in part
by the NSFC grant 10525314, 10503010, 10373001, 10773001, by
the CAS grant KJCX3-SYW-N2, MoST 973 grant 2007CB815401, and by NASA ATP
grant NNG06GI09G. Computing resources were in part provided
by the NASA High-End Computing (HEC) Program through the NASA
Advanced Supercomputing (NAS) Division at Ames Research Center.

\end{document}